# Seismic metamaterial: how to shake friends and influence waves ?

**Authors:** S. Brûlé[1]*, E.H. Javelaud[1], S. Enoch[2], S. Guenneau[2]*

**Affiliations:**

[1]Ménard, 91 620 Nozay, France.

[2]Aix Marseille Université, CNRS, Ecole Centrale Marseille, Institut Fresnel, 13013 Marseille, France.

*Correspondence to: stephane.brule@menard-mail.fr , sebastien.guenneau@fresnel.fr.

**Abstract:** Materials engineered at the micro- and nano-meter scale have had a tremendous and lasting impact in photonics and phononics, with applications ranging from periodic structures disallowing light and sound propagation at stop band frequencies, to subwavelength focussing and cloaking with metamaterials. Here, we present the description of a seismic test held on a soil structured at the meter scale using vibrocompaction probes. The most simplistic way to interact with a seismic wave is to modify the global properties of the medium, acting on the soil density and then on the wave velocity. The main concept is then to reduce the amplification of seismic waves at the free surface, called site effects in earthquake engineering. However, an alternative way to counteract the seismic signal is by modifying the distribution of seismic energy thanks to a metamaterial made of a grid of vertical, cylindrical and empty inclusions bored in the initial soil, in agreement with numerical simulations using an approximate plate model.

**One Sentence Summary:** We describe a seismic test held on a soil structured at the meter scale using vibrocompaction probes.

**Main Text:** In 1987, the groups of E. Yablonovitch and S. John reported the discovery of stop band structures for light *(1, 2)*. Photonic crystals (PCs) have since then held their promises with numerous applications ranging from nearly perfect mirrors for incident waves whose frequencies fall within stop bands of the PCs, to high-q cavities for PCs with structural defects *(3)*. The occurrence of stop bands in PCs also leads to anomalous dispersion whereby dispersion curves have a negative or vanishing group velocity. Dynamic artificial anisotropy, also known as all-angle-negative-refraction *(4-7)*, allows for focusing effects through a PC, as envisioned 45 years ago by V. Veselago *(8)*. With the advent of electromagnetic metamaterials *(9,10)* , J. Pendry pointed out that the image through the V. Veselago lens can be deeply subwavelength *(11)*, and exciting effects such as simultaneously negative phase and group velocity of light *(12)*, invisibility cloak *(13)* and tailored radiation phase pattern in epsilon near zero metamaterials were demonstrated *(14,15)*.

In parallel, research papers in phononic crystals provided numerical and experimental evidence of filtering *(16)* and focusing properties *(17)* of acoustic waves. Localized resonant structures for elastic waves propagating within three-dimensional cubic arrays of thin coated spheres *(18)* and fluid filled Helmholtz resonators *(19)* paved the way towards acoustic analogues of electromagnetic metamaterials *(18-21)*, including elastic cloaks *(22, 23,24)*. Control of elastic wave trajectories in thin plates was reported numerically *(25)* and experimentally *(26)* earlier this year, which prompted civil engineers at the Ménard company

*(27)* to explore potential routes towards metamaterials for surface seismic waves *(29, 32)* in civil engineering applications. In fact, Rayleigh waves are generated by anthropic sources such as an explosion or a tool impact or vibration (sledge-hammer, pile driving operations, vibrating machine footing, dynamic compaction, etc.). In 1968, R.D. Woods realized in situ tests with a 200 to 350 Hz source to show the effectiveness of isolating circular or linear empty trenches *(30)*. With the same geometry, these results were compared in 1988 with numerical modeling studies provided by P.K. Banerjee *(31)*. The main idea of this Report is to point out the possibilty to realize seismic metamaterials not only for high frequency anthropic sources but for the earthquakes' frequency range i.e. 0.1 to 50 Hz.

We first have to answer this fundamental question: What is the range of wavelengths taken into account for seismic building design? The amplification of seismic waves at the free surface, namely « site effects » may strengthen the impact of an earthquake in specific areas (e.g. Mexico in 1985). Indeed, when seismic waves propagate through alluvial layers or scatter on strong topographic irregularities, refraction/scattering phenomena may strongly increase the amplitude of the ground motion. It is then possible to observe stronger motions far away from the epicenter. At the scale of an alluvial basin, regarding buildings, seismic effects involve various phenomena as wave trapping, resonance of the whole basin, propagation in heterogeneous media, generation of surface waves on the basin edges *(28, 29)*. Structure damages due to seismic excitation is often directly correlated to local site condition in the form of motion amplification and/or soil liquefaction inducing ground deformation. An important fact is the low value of surface wave velocity, generated by natural seismic source or construction work activities, in superficial and under-consolidated recent material: less than 100 m/s to 300 m/s. In these so-called geomaterials, considering the 0.1 to 50 Hz frequency range, wavelengths of induced surface waves are shorter than direct P (Primary i.e. longitudinal compressional) and S (Secondary i.e. transverse shear) waves ones: from few meters to few hundreds of meters. This order of length is similar to that of buildings. This is the reason why we can expect building's resonance phenomena with some soil in case of earthquake. Indeed, the bottom line of concept of seismic metamaterials is that their size could be similar to that of the building project.

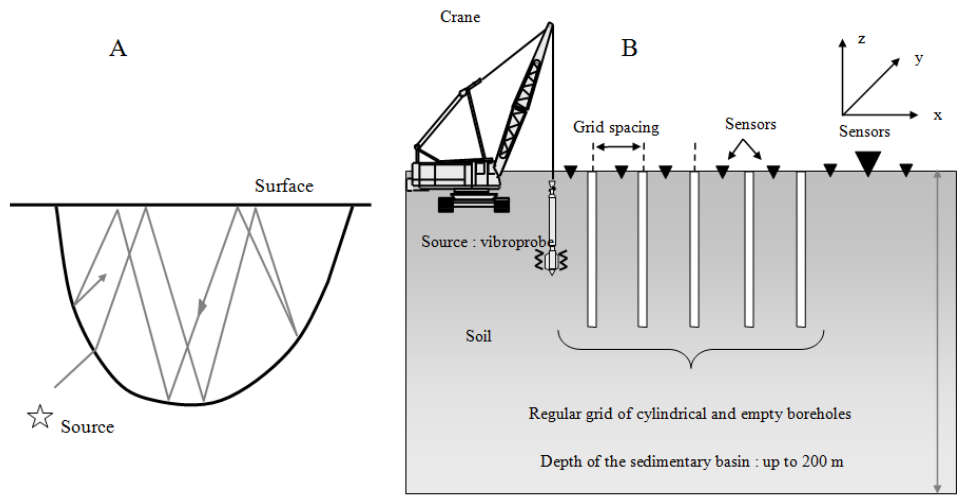

**Fig. 1.** Schematics of (A) Seismic wave in alluvium basin ; (B) Seismic testing device cross section in the xz-plane (see Fig. S1-2 in *(33)* for a photograph of the experiment).

A test zone consisting of a regular mesh of vertical cylindrical voids was carried out near the alpine city of Grenoble (France) in August 2012 (Fig. 1). The preliminary objective of this seismic field test is to point out analogies with control of electromagnetic *(10, 14, 15)* and acoustic *(16, 17)* waves by a quantitative approach. In theory, it seems realistic to influence seismic waves passing through an artificial anisotropic medium. However, soils possess particular characteristics: non elastic behavior, high rate of signal attenuation, large-scale heterogeneity, etc. These various uncertainties and the objective of realistic values for modeling require in situ tests to adjust soil's parameters as shear modulus, quality factor, etc. The measurement of the velocity of Rayleigh waves $V_R$ is given by a preliminary seismic test, pointing the wave time arrival at various offsets from the source, see Fig. S1-1 in *(33)*. We obtained $V_R$ = 78 m/s. The tested soil is a homogeneous silty clay. The thickness of the basin with similar deposits is up to 200 m. The length of columns is about 5 m and the grid spacing is smaller than 2 m: 1.73 m. The columns' mean diameter is 320 mm. A numerical simulation with finite elements shown in Fig. 2 predicts a stop band for elastic surface waves around 50 Hertz.

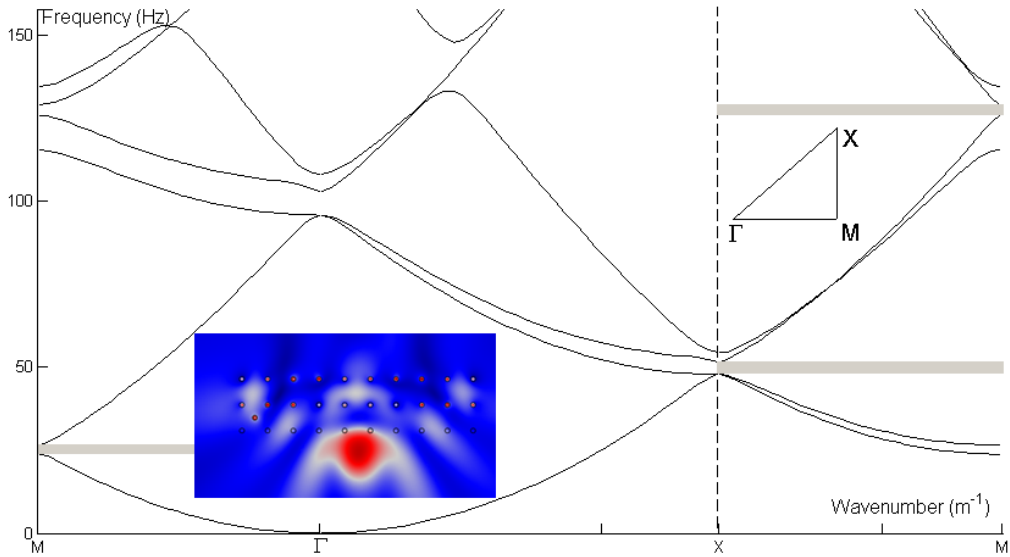

**Fig.2.** Simulated dispersion curves (frequency versus Bloch wavenumber describing the reduced Brillouin zone of vertices G=(0,0), X=($\pi$/d, $\pi$/d), M=($\pi$/d,0)) for a periodic plate of pitch d=1.73m and thickness 5m, with inclusions of diameter 0.32m and density 1/100[th] that of the surrounding medium (soil). The inset shows the plot of flexural wave (i.e. displacement in xz-plane in Fig. 1(b)) intensity for a forcing at 50 Hz (frequency in the second partial stop band along XM), which is located as in Fig. 3.

The frequency of the vibrating source in the experiments, 50 Hz, with 14 mm of lateral amplitude in xy-plane, should therefore lead to very strong reflection of surface elastic waves by the large scale metamaterial (see inset in Fig. 2). The experimental grid is made of three discontinuous lines of ten boreholes 320 mm in diameter. Sensors are three components velocimeters (x, y, z) with a corner frequency of 4.5 Hz (-3dB at 4.5Hz) electronically corrected to 1 Hz. Twenty sensors were used simultaneously with a common time-base. In order to map completely the energy's field, the sensors were set four times on site before and after carrying out the boreholes.

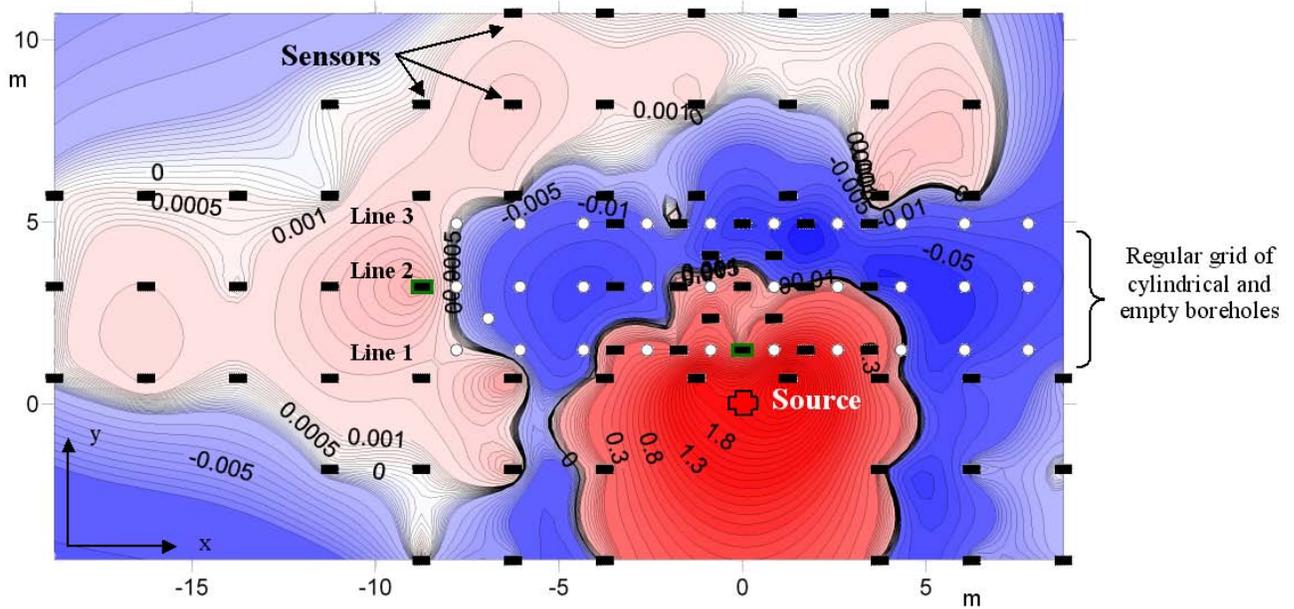

**Fig. 3**. Experimental results map after interpolation between sensors: Difference of the measured energy's field (arbitrary units) after and before carrying out the boreholes. Black rectangles symbolise sensors.

The sensors A1 and B1 were kept at the same position during the whole experiment. They are the two sensors in Figure 3 that have green frames: A1 is the sensor right above the source; B1 is just on the left side of the boreholes. Due to the strong soil attenuation, the probe is close to the grid (1.5 m). Each record is about two minutes long and can be divided into three parts: a pre-experiment part, record of ambient vibration noise, followed by the probe starting phase during which the frequency of the generated signal increases up to 50 Hz and the probe is set in place; the experimental phase itself, about one minute long and a post-experiment part, during which the mechanical probe stops, and which end by ambient vibration noise. For each grid, the effective experimental phase's length is determined for the A1 fixed sensor. It is defined as the aforementioned experimental length reduced by ten seconds at both ends to ensure that any effect of the probe's starting and ending are well removed. The signals' average energy per second is then computed for each sensor. They are consecutively normalized by the A1's energy per second, so as to reconstruct a uniform energy field over the whole experimental area. The boreholes' effect on the energy field is given Fig. 3 where the difference of the measured energy's field after and before carrying out the boreholes is displayed with interpolation between sensors.

**Experimental result discussion:** According to the 50 Hz source, the wavelength of the signal propagated at the source is 1.56 m compared to the 1.73 m separating each borehole's center. The center of the vibroprobe is located at 1.5 m from the first line of the grid, in the perpendicular direction. In this soft soil, the attenuation may be significant. We verify that the signal to noise ratio remains up to two at 10 m from the source. Fig. 3 shows a concentration of energy in a 10m wide zone strongly attenuated towards the structure centered around the vibroprobe, in good agreement with the inset in Fig. 2. In this area and near the source, the difference between the measured energy's field after and before carrying out the boreholes can reach twice the initial value confirming a strong reflection of surface waves by the seismic metamaterial. The signal hardly exceeds the second row of boreholes in Fig. 3 showing the efficiency of this device, for this geometry and a 50 Hz source in soft soils.

**References and Notes:**

1. E. Yablonovitch, Inhibited spontaneous emission in solid-state physics and electronics. *Phys. Rev. Lett.* **58**, 2059-2062 (1987).
2. S. John, Strong localization of photons in certain disordered dielectric superlattices. *Phys. Rev. Lett.* **58**, 2486-2489 (1987).
3. K. Srinivasan, O. Painter, Momentum space design of high-Q photonic crystal optical cavities. *Opt. Express* **10**, 670-684 (2002).
4. R. Zengerle, Light propagation in singly and doubly periodic waveguides. *J. Mod. Opt.* **34**, 1589-1617 (1987).
5. M. Notomi, Theory of light propagation in strongly modulated photonic crystals: Refractionlike behavior in the vicinity of the photonic band gap. *Phys. Rev. B* **62**, 10696-10705 (2000).
6. B. Gralak, S. Enoch, G. Tayeb, Anomalous refractive properties of photonic crystals. *J. Opt. Soc. Am.* **A 17**, 1012-1020 (2000).
7. C. Luo, S. G. Johnson, J. D. Joannopoulos, J. B. Pendry, All-angle negative refraction without negative effective index. *Phys. Rev. B* **65**, 201104 (2002).
8. V. G. Veselago, The electrodynamics of substances with simultaneously negative values of ε and µ. *Soviet Physics Uspekhi* **10**, 509-514 (1968).
9. J. B. Pendry, A. J. Holden, D. J. Robbins, W. J. Stewart, Magnetism from conductors and enhanced nonlinear phenomena. *IEEE Transactions on Microwave Theory and Techniques* **47**, 2075-2084 (1999).
10. D. R. Smith, W. J. Padilla, V. C. Vier, S. C. Nemat-Nasser, S. Schultz, Composite medium with simultaneously negative permeability and permittivity. *Phys. Rev. Lett.* **84**, 4184-4187 (2000).
11. J. B. Pendry, Negative refraction makes a perfect lens. *Phys. Rev. Lett.* **85**, 3966-3969 (2000).
12. G. Dolling, C. Enkrich, M. Wegener, C. M. Soukoulis, S. Linden, Observation of simultaneous negative phase and group velocity of light in a metamaterial. *Science* **312**, 892-894 (2006).
13. D. Schurig *et al.*, Metamaterial electromagnetic cloak at microwave frequencies. *Science* **314**, 977-980 (2006).
14. S. Enoch, G. Tayeb, P. Sabouroux, N. Guérin, P. Vincent, A metamaterial for directive emission. *Phys. Rev. Lett.* **89**, 213902 (2002).
15. A. Alu, M. G. Silveirinha, A. Salendrino, N. Engheta, Epsilon-near-zero metamaterials and electromagnetic sources: Tailoring the radiation phase pattern. *Phys. Rev. B* **75**, 155410 (2007).
16. R. Martinez-Sala *et al.*, Sound attenuation by sculpture. *Nature* **378**, 241 (1995).
17. A. Sukhovich *et al.*, Experimental and theoretical evidence for subwavelength imaging in phononic crystals. *Phys. Rev. Lett.* **102**, 154301 (2009).
18. Z. Liu *et al.*, Locally resonant sonic materials. *Science* **289**, 1734-1736 (2000).

19. N. Fang *et al.*, Ultrasonic metamaterial with negative modulus. *Nat. Mater.* **5**, 452-456 (2006).
20. J. Christensen, F. J. García de Abajo, Anisotropic metamaterials for full control of acoustic waves. *Phys. Rev. Lett.* **108**, 124301 (2012).
21. R. V. Craster, S. Guenneau, *Acoustic Metamaterials: Negative Refraction, Imaging, Lensing and Cloaking* (Springer-Verlag, Springer Series in Materials Science, January 2013).
22. G. W. Milton, M. Briane, J. R. Willis, On cloaking for elasticity and physical equations with a transformation invariant form. *New J. Phys*. **8,** 248 (2006).
23. M. Brun, S. Guenneau, A. B. Movchan, Achieving control of in-plane elastic waves. *Appl. Phys. Lett.* **94**, 061903 (2009).
24. A. N. Norris, A. L. Shuvalov, Elastic cloaking theory. *Wave Motion* **48**, 525-538 (2011).
25. M. Farhat, S. Guenneau, S. Enoch, Broadband cloaking of bending waves via homogenization of multiply perforated radially symmetric and isotropic thin elastic plates. *Phys. Rev. B* **85**, 020301 R (2012).
26. N. Stenger, M. Wilhelm, M. Wegener, Experiments on elastic cloaking in thin plates. *Phys. Rev. Lett.* **108**, 014301 (2012).
27. http://www.menard-web.com/internetmenard.nsf/HTML/home_en.html.
28. J.F. Semblat and A. Pecker, *Waves and vibrations in soils: earthquakes, traffic, shocks, construction works*, (IUSS Press, Pavia, 2009).
29. S. Brûlé, E. Javelaud, Could deep soil densification impact the seismic site effect? Paper presented at the 9th Annual International Conference on Urban Earthquake Engineering 9th CUEE, Tokyo, **02-281**, 497-501 (2012).
30. R.D. Woods, "Screening of surface waves in soils" (Tech. Rep. IP-804, University of Michigan, 1968).
31. P.K. Banerjee, S. Ahmad, K.Chen, Advanced application of BEM to wave barriers in multi-layered three-dimensional soil media. Earthquake Eng. & Structural Dynamics **16**, 1041-1060 (1988).
32. S. Brûlé, E. Javelaud, S. Guenneau, S. Enoch, D. Komatitsch, Seismic metamaterials, Paper presented at the 9th International Conference of the Association for Electrical, Transport and Optical Properties of Inhomogeneous Media (2012).
33. Supplementary Materials www.sciencemag.org

**Acknowledgments:** S.G. is thankful for European funding through ERC Starting Grant ANAMORPHISM. S.B. thanks the MENARD's Earthquake Engineering Team.

## Supplementary materials:

The seismic metamaterial test is really challenging to model in full (3D Navier equations in unbounded heterogeneous media). We therefore opted for an asymptotic model, which only captures the wave physics at the air-soil interface. We do not claim to have a complete understanding of the elastic wave propagation. However, there is a qualitatively good agreement between our numerical simulations, in Fig. 2, and Ménard's company experiments, in Fig. 3. Note that the seismic source frequency at 50 Hz in Fig. 3 is located inside the first partial stop band in Fig. 2, which confirms the stop band origin of the wave reflection.

We consider the following approximate plate model for surface flexural waves

$$\rho^{-1}\nabla.(E^{1/2}\nabla\rho^{-1}\nabla.(E^{1/2}\nabla\Psi))-\beta^4\Psi=0$$

where $\Psi(x,y)$ is the amplitude of displacement along the z-axis in Fig. 1b, and $\rho$ is the heterogeneous density of the plate: in the soil $\rho=1500$ kg/m$^3$ and in bored holes $\rho=1.2$ kg/m$^3$. Moreover, the spectral parameter $\beta^4=\omega^2\rho h/D$, with $\omega$ the angular flexural wave frequency, h the plate thickness (assumed to be 5m), and D the plate rigidity, which in the present case is $D=Eh^3/(12(1-\nu^2))$. From the Ménard company characterization of the soil, we chose E=100 MPA and $\nu=0.3$. In the finite element model solved with the COMSOL MULTIPHYSICS software, we set Floquet-Bloch boundary conditions on either sides of a periodic cell (in the xy-plane) of sidelength d=1.73m in order to compute the dispersion diagram shown in Fig. 2. We also set perfectly matched layers on either boundaries of the domain in order to account for the unbounded domain in the xy-plane for the simulation shown in the inset of Fig. 2.

Regarding the experimental results shown in Fig. 3, some word would be in order for the experimental setup. Twenty sensors were used simultaneously with a common time-base. In order to map completely the energy's field, the sensors were set four times on site (green, blue, pink and orange grids in Fig. 2) before and after carrying out the boreholes (Fig. S1-1 and S1-2). The fact that the elastic energy is 2.3 times larger at the source point when it is in presence of the metamaterial in Fig. 3 is reminiscent of the Local Density of States obtained for a source placed near a mirror in optics.

**Fig. S1-1**. Seismic test metamaterial map with the borehole pattern and four set of velocimeters (green, blue, pink and orange grids).

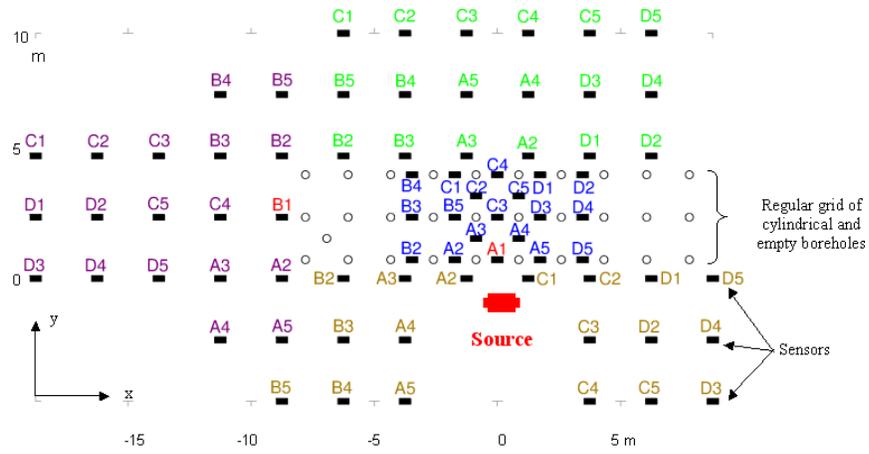

**Fig. S1-2**. Photograph of the seismic metamaterial experiment from the Ménard company.

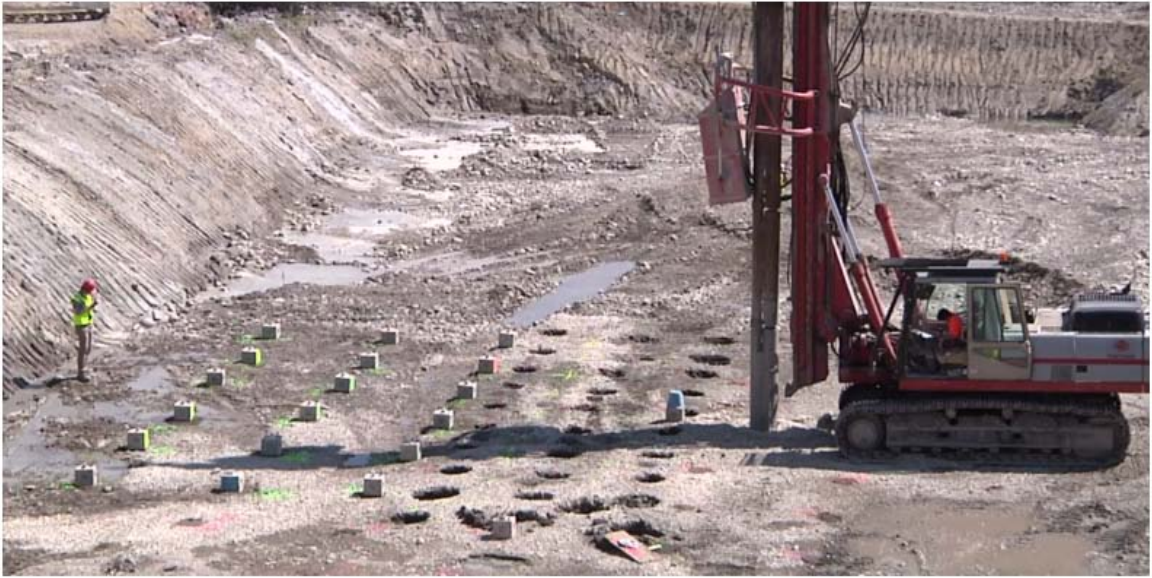

Sensitive three components Velocimeters

Five meters deep 320 mm holes

Source: - frequency: 50 Hz
         - horizontal displacement : 14 mm